\documentclass[conference]{IEEEtran}
\usepackage[T1]{fontenc}
\IEEEoverridecommandlockouts
\usepackage{cite,amsmath,amssymb,amsfonts,subcaption}
\usepackage[colorlinks=true, urlcolor=black]{hyperref}
\usepackage{algorithmic,graphicx,textcomp,xcolor}
\usepackage{booktabs, multirow,colortbl}
\usepackage[T1]{fontenc}
\usepackage{tikz}
\def\BibTeX{{\rm B\kern-.05em{\sc i\kern-.025em b}\kern-.08em
    T\kern-.1667em\lower.7ex\hbox{E}\kern-.125emX}}
\begin{document}

\title{Beyond Forced Modality Balance: Intrinsic Information Budgets for Multimodal Learning}
\author{\IEEEauthorblockN{
Zechang Xiong\IEEEauthorrefmark{1}$^1$\thanks{\IEEEauthorrefmark{1} Contributed equally.},
Da Li\IEEEauthorrefmark{1}$^2$,
Kexin Tang\IEEEauthorrefmark{2}$^1$\thanks{\IEEEauthorrefmark{2} Project leader.}, 
Pengyuan Li$^3$, 
Wenkang Kong$^3$, 
Yulan Hu\IEEEauthorrefmark{3}$^1$\thanks{\IEEEauthorrefmark{3} Corresponding author.}
}
\IEEEauthorblockA{
$^1$ Amap, Alibaba Group, Beijing, China\\
$^2$ Institute of Computing Technology, Chinese Academy of Sciences, Beijing, China\\
$^3$ Beijing Jiaotong University, Beijing, China\\
\{jerryhusing,\,lida.ucas\}@gmail.com,\, jixuan.tkx@alibaba-inc.com,\, \\
\{pengyuanli,\,wenkang.kong\}@bjtu.edu.cn,\,huyulan@alibaba-inc.com}
}
\maketitle

\begin{abstract}
Multimodal models often converge to a dominant-modality solution, in which a stronger, faster-converging modality overshadows weaker ones. This modality imbalance causes suboptimal performance.
Existing methods attempt to balance different modalities by reweighting gradients or losses. However, they overlook the fact that each modality has finite information capacity.
In this work, we propose \textbf{IIBalance}, a multimodal learning framework that aligns the modality contributions with Intrinsic Information Budgets (IIB).
We propose a task-grounded estimator of each modality’s IIB, transforming its capacity into a global prior over modality contributions. 
Anchored by the highest-budget modality, we design a prototype-based relative alignment mechanism that corrects semantic drift only when weaker modalities deviate from their budgeted potential, rather than forcing imitation. 
During inference, we propose a probabilistic gating module that integrates the global budgets with sample-level uncertainty to generate calibrated fusion weights. 
Experiments on three representative benchmarks demonstrate that IIBalance consistently outperforms state-of-the-art balancing methods and achieves better utilization of complementary modality cues. Our code is available at: \href{https://github.com/XiongZechang/IIBalance}{https://github.com/XiongZechang/IIBalance}.
\end{abstract}


\section{Introduction}
Multimodal learning is expected to exploit complementary information across modalities. \cite{wang2020what}. When one modality becomes unreliable due to noise or adverse environmental conditions, the others are expected to provide complementary evidence, enabling multimodal models to achieve robust perception in complex scenarios. However, modern end-to-end training fails to achieve this ideal state~\cite{peng2022balanced, wu2022characterizing}. Empirical studies reveal that multimodal models suffer from modality imbalance, where a dominant modality converges faster and suppresses the learning of weaker modalities, causing the model to overrely on a single stream and underutilize complementary cues \cite{fan2023pmr}. Such behaviors are consistent with shortcut learning effects observed in unimodal networks \cite{geirhos2020shortcut} and result in impaired generalization performance under distribution shifts or when modalities are absent~\cite{lee2023multimodal}.

\begin{figure}[t]
  \centering
  \includegraphics[width=0.8\linewidth]{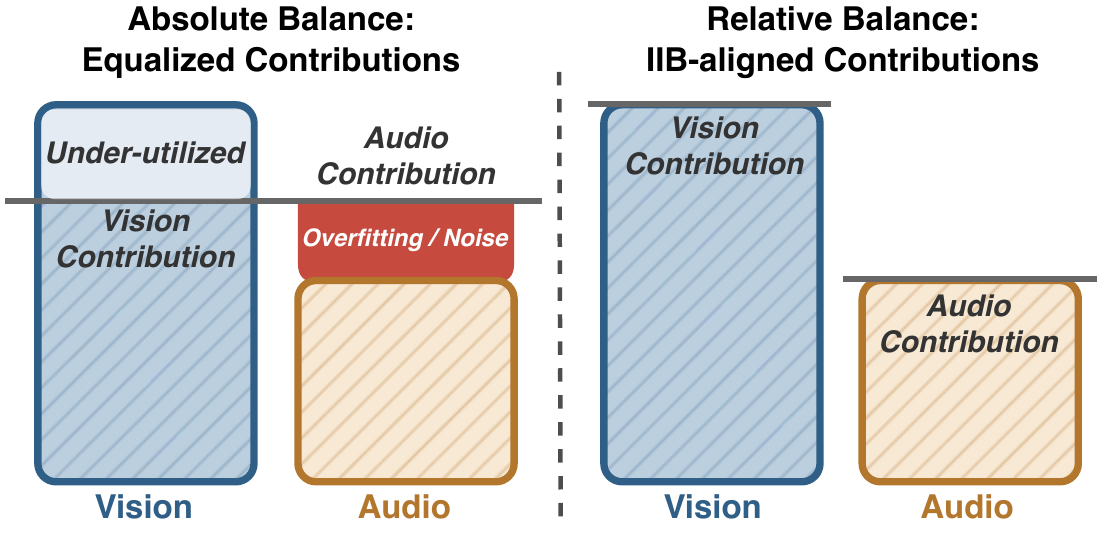}
  \caption{\textbf{Absolute vs.\ relative modality balancing.}
  \textbf{Left:} absolute balance enforces equal contribution across modalities, which can under-utilize high-capacity modalities while pushing low-capacity modalities to overfit residual noise.
  \textbf{Right:} Relative balance allocates each modality’s contribution in accordance with its \textbf{Intrinsic Information Budget (IIB)}, encouraging capacity-aware utilization.}
  \label{fig:intro}
\end{figure}

To alleviate these problems, existing balancing methods mainly encourage modalities to learn at similar rates or contribute comparably through gradient, loss, or fusion reweighting~\cite{peng2022balanced,fan2023pmr}. First, the discriminative information carried by different modalities is inherently unequal. In many perception tasks, visual signals provide richer, more directly task-relevant cues than audio or auxiliary streams. Enforcing equal contributions from different modalities may result in multimodal models relying on low-information-density modalities to fit residuals not covered by high-information-density modalities, which could introduce interference instead of achieving beneficial modal complementarity~\cite{wei2024mmpareto}. As illustrated in Fig.~\ref{fig:intro}, enforcing absolute balance tends to suppress high-capacity modalities and push low-capacity modalities to fit residual noise.
In addition, the information differences between modalities are sample-specific. For example, visual signals may degrade in low light, while audio can become unreliable in noisy scenes, and a single global balancing rule cannot capture this dynamic inversion of modality reliability~\cite{han2023trusted}.

These observations suggest that multimodal learning should shift from a pursuit of absolute balance between modalities to a relative balance. It is essential to consider the information-carrying capacity of different modalities and employ flexible fusion strategies to enable multimodal models to achieve optimal performance across diverse scenarios. In this view, an effective balance mechanism should not force modalities to contribute equally. Instead, it should align each modality’s contribution with its intrinsic information quality, modeled as a modality-specific prior, while still allowing adaptive, sample-specific reweighting~\cite{peng2022balanced}.

Motivated by these insights, we propose \textbf{IIBalance}, a budget-aware balancing principle that treats modality contributions as bounded resources, so that weak modalities are guided to saturate their own capacity instead of imitating the dominant modality. 
Specifically, we explicitly formalize relative balancing through an Intrinsic Information Budget (IIB), which provides a dataset-level prior that quantifies task-relevant information each modality can reliably contribute.
Building on the IIB, we design a prototype-guided relative alignment mechanism that enables the model to adaptively learn from different modalities, in which the strong modality serves as an adaptive reference that guides weaker modalities only when they underuse their estimated budget. This limits overfitting from excessive cross-modal supervision.
From a Bayesian-inspired viewpoint, we propose an uncertainty-aware dynamic fusion strategy that treats IIB as a prior and combines it with sample-level predictive uncertainty as a likelihood proxy to obtain calibrated fusion weights. This prior–likelihood formulation is consistent with uncertainty-aware multimodal fusion \cite{subedar2019uncertainty,tian2020uno}. The resulting fusion weights adhere to the IIB prior but are adjusted by sample-level uncertainty, up-weighting low-uncertainty modalities, down-weighting uncertain ones, and reverting to the prior when the evidence becomes unreliable.

Our contributions can be summarized as follows: 
\begin{enumerate}
    \item We introduce Intrinsic Information Budget, a dataset-level prior for capacity-aware relative modality balance.
    \item We develop an IIB-guided framework with prototype-guided alignment for training and uncertainty-aware adaptive fusion for inference.
    \item Extensive experiments on three audio-visual benchmarks validate the effectiveness of the proposed method.
\end{enumerate}

\section{Related Work}

\subsection{Uncertainty-Aware Fusion}
Uncertainty-aware fusion improves robustness by estimating modality reliability and adapting fusion weights \cite{subedar2019uncertainty}. Early methods handle unseen degradations by uncertainty-scaled predictions combined with probabilistic fusion rules \cite{tian2020uno}. More recent approaches perform confidence-aware, sample-adaptive fusion via gating or dynamic selection, aiming to trust the informative modality/features for each instance \cite{han2022mmdynamics, zhang2023qmf}. Our method follows the uncertainty-aware fusion principle but further constrains reliability weighting with a global, modality-specific capacity prior, ensuring that instance-wise reweighting does not demand more from a weak modality than it can stably contribute.

\subsection{Imbalanced Multimodal Learning}
Recent studies have identified modality competition under optimization imbalance as a key source of multimodal performance gaps, as dominant modalities learn faster and gradually suppress weaker ones \cite{wang2020what,pmlr-v162-huang22e}. Existing methods can be roughly grouped by how they mitigate imbalance. The first group directly modifies optimization signals, e.g., through gradient-level modulation to boost under-optimized modalities \cite{peng2022balanced,li2023agm} or objective shaping to reduce representation bias in fine-grained settings \cite{xu2023mmcosine}. The second group changes the training paradigm to reduce cross-modal interference, for instance, by alternating unimodal adaptation while maintaining shared cross-modal knowledge \cite{zhang2024mla}. The third group targets fine-grained imbalance by estimating per-sample modality contribution and enhancing low-contributing modalities \cite{wei2024valuation, fan2023pmr}. In contrast to purely dynamics-driven rebalancing, we emphasize respecting each modality's inherent capacity and avoiding forced equal contribution when a modality cannot reliably support it.

\section{Preliminaries}
\label{sec:prelim}

\subsection{Multimodal Learning Formalization}

We consider a supervised multimodal classification with $M$ input modalities. The training set is
$\mathcal{D}=\{(x_1^{(i)},\dots,x_M^{(i)},y^{(i)})\}_{i=1}^N$, where $x_m^{(i)}\in\mathcal{X}_m$ denotes the input of modality $m$ and $y^{(i)}\in\{1,\dots,C\}$ is the label. For each modality $m$, an encoder $g_m:\mathcal{X}_m\!\to\!\mathbb{R}^d$ produces a representation $z_m^{(i)}=g_m(x_m^{(i)})$, and a unimodal classifier $f_m:\mathbb{R}^d\!\to\!\Delta^{C-1}$ outputs $p_m(\cdot\mid x_m^{(i)})=f_m(z_m^{(i)})$.
In end-to-end training, gradients from the fused loss are backpropagated through all modality encoders simultaneously, which often leads to a \textbf{\textit{dominant-modality}} solution, where one modality with faster convergence or stronger signal overwhelms the others and prevents weaker modalities from learning informative representations. Our goal is to design a training scheme that respects the intrinsic capacity of each modality while still exploiting their complementarity.

\subsection{Intrinsic Information Budget}
Different modalities often have asymmetric capacities and noise levels; forcing equal contribution may amplify unreliable signals and impact the performance of downstream tasks. Modal balance is meaningful only when each modality provides a comparable amount of task-related information to downstream tasks. Therefore, we introduce a dataset-level prior, called the \textbf{{Intrinsic Information Budget (IIB)}}, to quantify each modality's reliable information capacity and use it as a reference for relative balancing. To estimate each modality's intrinsic task-relevant capacity, we compute the IIB prior in a straightforward and verifiable manner. For each modality $m \in \{1,\dots,M\}$, we first train the unimodal encoder--classifier pair $(g_m, f_m)$ on modality-specific inputs $x_m^{(i)}$ until convergence. Let $p_m(\cdot \mid x_m^{(i)}) = f_m(g_m(x_m^{(i)})) \in \Delta^{C-1}$ denote the predictive distribution over $C$ classes for sample $(x_1^{(i)}, \dots, x_M^{(i)}, y^{(i)})$. We use the Shannon entropy normalized by $\log C$, $\mathcal{H}(p) = - \frac{1}{\log C} \sum_{c=1}^C p_c \log p_c \in [0,1]$, as a measure of predictive uncertainty.
\begin{figure*}[htbp!]
    \centering
    \includegraphics[width=0.9\linewidth]{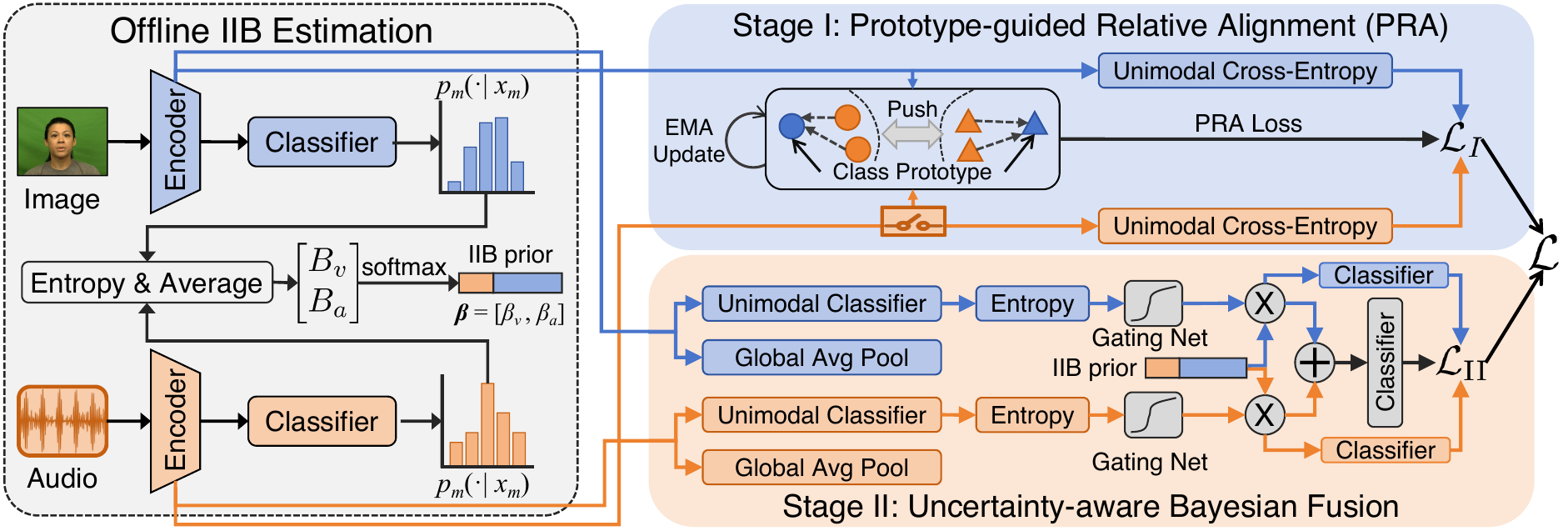}
    \caption{
    Overview of \textbf{IIBalance}. An intrinsic information budget prior is estimated from unimodal prediction entropy. \textbf{\textit{Stage I}} learns unimodal features with prototype-guided relative alignment. \textbf{\textit{Stage II}} conducts uncertainty-aware Bayesian fusion using entropy-conditioned gating. Here, $\otimes$ denotes element-wise multiplication and $\oplus$ denotes element-wise addition.
    }
    \label{fig:model_overview}
\end{figure*}
We then define a normalized signal-confidence proxy $B_m$ as the dataset-level expected negative entropy:
\begin{equation}
B_m \;=\; \mathbb{E}_{(x_1,\dots,x_M,y)\sim\mathcal{D}}\big[\,1 - \mathcal{H}\big(p_m(\cdot \mid x_m)\big)\,\big],
\end{equation}
which in practice is approximated by the empirical average over training samples. Intuitively, a larger $B_m$ indicates that modality $m$ produces more confident predictions on average and thus possesses higher intrinsic discriminative capacity.

We convert $\{B_m\}$ into a \textit{relative budget} prior via a softmax-style normalization with temperature $\tau > 0$:
\begin{equation}
\beta_m \;=\; \frac{\exp\!\big(B_m/\tau\big)}{\sum_{k=1}^M \exp\!\big(B_k/\tau\big)}, 
\qquad m=1,\dots,M.
\end{equation}
The vector $\boldsymbol{\beta} = [\beta_1,\dots,\beta_M]$ serves as a dataset-level prior over modality contributions and will be used in both the alignment regularization and the fusion stage. We refer to the \textit{anchor modality} as the modality with the highest budget, that is $m^* = \arg\max_m \beta_m$. In practice, we estimate $\{B_m\}$ and $\boldsymbol{\beta}$ once from the pretrained unimodal classifiers and keep this IIB prior fixed during subsequent multimodal training.

\section{Method}
\subsection{Overview}
Given a training sample $(x_1^{(i)},\dots,x_M^{(i)},y^{(i)})$, we first obtain the unimodal representation $z_m^{(i)} = g_m(x_m^{(i)})$ and the predictive distribution $p_m(\cdot \mid x_m^{(i)}) = f_m(z_m^{(i)})$ as defined in Section~\ref{sec:prelim}. As shown in Fig.~\ref{fig:model_overview}, IIBalance consists of two coupled stages to normalize the outputs of these encoders and compute sample-adaptive fusion weights.
\textbf{\textit{Stage I}} focuses on relative alignment in representation. We select an anchor modality $m^*$ according to the IIB prior $\boldsymbol{\beta}$, maintain class prototypes from this anchor modality, and align weaker modalities to the anchor prototypes using a prototype-based contrastive loss whose strength is controlled by the budget gap $\beta_{m^*} - \beta_m$. \textbf{\textit{Stage II}} performs uncertainty-aware Bayesian fusion. We design a lightweight gating network that, for each sample, fuses the global prior $\boldsymbol{\beta}$ with sample-level signals derived from per-modality predictive uncertainty and pooled features to produce normalized fusion weights. The fused representation is used to make the final prediction. In the inference phase, we use the learned encoders and gating network to compute fusion weights and obtain predictions from the fused branch.

\subsection{Prototype-guided Relative Alignment}
Rather than forcing weaker modalities to mimic representation of stronger modalities, we encourage weaker modalities to align with class prototypes computed from the anchor modality. This preserves each modality's unique characteristics while correcting semantic drift.

\paragraph{Anchor prototypes}
For the anchor modality $m^*$, we maintain a set of class prototypes $\{P_c^{m^*}\}_{c=1}^C$, with $P_c^{m^*} \in \mathbb{R}^d$, updated online by an exponential moving average (EMA). Let $\mathcal{B}$ denote the current mini-batch and $\mathcal{B}_c = \{i \in \mathcal{B} : y^{(i)} = c\}$ the indices of samples from class $c$. The prototype is updated in the following manner:
\begin{equation}
P_c^{m^*} \leftarrow \rho\, P_c^{m^*} + (1-\rho)\, \mathrm{Mean}\big(\{z_{m^*}^{(i)} : i \in \mathcal{B}_c\}\big),
\end{equation}
where $\rho \in [0,1)$ is the EMA momentum and the mean is taken over anchor-modality features in the current mini-batch (or an optional memory buffer). Both $P_c^{m^*}$ and $z_{m^*}^{(i)}$ lie in $\mathbb{R}^d$, so their inner product is well defined.

\paragraph{Prototype-guided Relative Alignment (PRA)}
For a non-anchor modality $m \neq m^*$, let $z_m^{(i)} \in \mathbb{R}^d$ be the modality feature for the $i$-th sample and $y^{(i)}$ its label. We define a prototype contrastive loss that pulls $z_m^{(i)}$ towards its class prototype $P_{y^{(i)}}^{m^*}$ and pushes it away from other class prototypes:
\begin{equation}\label{eq:pra}
\mathcal{L}_{\mathrm{PRA}}^{m} \;=\; -\frac{1}{|\mathcal{B}|}\sum_{i \in \mathcal{B}} \log 
\frac{\exp\!\big( z_m^{(i)}\!\cdot\! P_{y^{(i)}}^{m^*} \!/\! \tau_p\big)}
{\sum_{c=1}^C \exp\!\big( z_m^{(i)}\!\cdot\! P_{c}^{m^*} \!/\! \tau_p\big)},
\end{equation}
where $\tau_p > 0$ is a temperature for prototype contrast. The dot products $z_m^{(i)}\!\cdot\! P_c^{m^*}$ are scalars, and the loss is the standard cross-entropy over the $C$ prototypes for modality $m$.

\paragraph{Budget-gap controlled alignment strength}
The alignment degree for a weak modality is determined by its budget gap relative to the anchor:
\begin{equation}
\lambda_m \;=\; \mathrm{ReLU}\big(\beta_{m^*} - \beta_m\big), \qquad m\neq m^*.
\end{equation}
Thus, the modalities with a budget close to the anchor's receive little alignment pressure, while those modalities with a low budget are more strongly encouraged to align towards class centers instead of raw anchor features.

\paragraph{Stage I Training objective}
\textit{Stage I} combines unimodal supervision with prototype-guided relative alignment for non-anchor modalities. For a mini-batch $\mathcal{B}$, the objective is formalized as:
\begin{equation}\label{eq:stageI}
\mathcal{L}_{\mathrm{I}} \;=\; 
\frac{1}{|\mathcal{B}|}\sum_{i \in \mathcal{B}} \sum_{m=1}^M 
\mathcal{L}_{\mathrm{CE}}\big(f_m(z_m^{(i)}), y^{(i)}\big)
\;+\; \sum_{m\neq m^*} \lambda_m \, \mathcal{L}_{\mathrm{PRA}}^{m},
\end{equation}
where $\mathcal{L}_{\mathrm{CE}}$ is the cross-entropy loss for the unimodal classifier $f_m$. The overall \textit{Stage~I} loss is a scalar, and the hyperparameters $\rho$, $\tau_p$, and $\lambda_m$ are fixed during training once the IIB prior $\boldsymbol{\beta}$ is computed.

\subsection{Uncertainty-aware Bayesian Fusion}
\textit{Stage~I} provides a global relative balancing by aligning modality contributions to the dataset-level IIB prior. However, multimodal reliability is often sample-specific: even a strong modality can be corrupted for some instances, and a weak modality can be informative for others. Thus, global alignment alone cannot determine how much each modality should be trusted per sample. Therefore, we introduce \textit{Stage~II} to compute sample-adaptive fusion weights that respect the global IIB prior while down-weighting unreliable modalities based on uncertainty.
After the unimodal encoders and anchor prototypes are regularized in \textit{Stage~I}, \textit{Stage~II} yields sample-adaptive fusion weights by integrating the global IIB prior $\boldsymbol{\beta}$ with sample-level signals based on predictive uncertainty and feature statistics.

\paragraph{Sample-level uncertainty and dynamic signals}
For the $i$-th sample with modality $m$, we measure predictive uncertainty by the normalized entropy:
\begin{equation}
u_{m}^{(i)} \;=\; \mathcal{H}\big(p_m(\cdot\mid x_m^{(i)})\big),
\end{equation}
where $\mathcal{H}(\cdot)$ is the entropy defined in Section~\ref{sec:prelim}. It captures how uncertain modality $m$ is about the $i$-th sample and reflects both data and model uncertainty. In addition, we extract simple pooled statistics from each modality feature, denoted $\mathrm{Pool}(z_m^{(i)}) \in \mathbb{R}^{d_p}$. We use global average pooling here.
We form the input to the gating network as:
\begin{equation}
\phi^{(i)} \;=\; \big[\,u_{1}^{(i)},\dots,u_{M}^{(i)};\;\mathrm{Pool}(z_{1}^{(i)}),\dots,\mathrm{Pool}(z_{M}^{(i)})\,\big],
\end{equation}
and let $G: \mathbb{R}^{d_g} \to \mathbb{R}^M$ be a lightweight network that maps $\phi^{(i)}$ to per-modality logits $G_m(\phi^{(i)})$.

\paragraph{Bayesian-inspired weight computation}
We compute an unnormalized fusion score for modality $m$ as:
\begin{equation}\label{eq:alpha}
\alpha_{m}^{(i)} \;=\; \underbrace{\beta_m}_{\text{prior}} \;\cdot\; 
\underbrace{\exp\!\big(-u_{m}^{(i)}\big)}_{\text{likelihood proxy}} \;\cdot\;
\underbrace{\sigma\!\big(G_m(\phi^{(i)})\big)}_{\text{calibration}},
\end{equation}
where $\sigma(\cdot)$ is the sigmoid activation. Here $\beta_m$ encodes the global intrinsic capacity of modality $m$, $\exp(-u_m^{(i)})$ down-weights locally uncertain evidence, and $\sigma(G_m(\phi^{(i)}))$ provides a learned, sample-dependent calibration factor. This factorization is Bayesian-inspired in the sense that it treats $\beta_m$ as a prior over modality reliability and the remaining terms as a data-dependent likelihood proxy.

We normalize $\{\alpha_{m}^{(i)}\}_{m=1}^M$ to obtain fusion weights and compute the fused representation as:
\begin{equation}
\tilde{w}_{m}^{(i)} \;=\; \frac{\alpha_{m}^{(i)}}{\sum_{k=1}^M \alpha_{k}^{(i)}}, \qquad
Z_{\mathrm{fused}}^{(i)} \;=\; \sum_{m=1}^M \tilde{w}_{m}^{(i)} \, z_m^{(i)}.
\end{equation}
Each $\tilde{w}_{m}^{(i)}$ is non-negative and sums to one across modalities, so $Z_{\mathrm{fused}}^{(i)}$ is a convex combination of modality features in $\mathbb{R}^d$.

\paragraph{Stage II Training objective}
The objective combines a cross-entropy term on the fused representation and a weighted unimodal auxiliary term that encourages unimodal correctness in proportion to their contribution. It is defined as:
\begin{equation}\label{eq:stageII}
\begin{aligned}
\mathcal{L}_{\mathrm{II}} \;=\; 
\frac{1}{|\mathcal{B}|}\sum_{i \in \mathcal{B}} 
\mathcal{L}_{\mathrm{CE}}\big(f_{\mathrm{fuse}}(Z_{\mathrm{fused}}^{(i)}), y^{(i)}\big)
\\[0.2em]
+ \;\gamma \,\frac{1}{|\mathcal{B}|}\sum_{i \in \mathcal{B}} \sum_{m=1}^M 
\tilde{w}_{m}^{(i)} \,\mathcal{L}_{\mathrm{CE}}\big(f_m(z_m^{(i)}), y^{(i)}\big),
\end{aligned}
\end{equation}
where $\mathcal{B}$ denotes a mini-batch, $\gamma \ge 0$ balances the auxiliary unimodal term and $f_{\mathrm{fuse}}(\cdot)$ is the fused classifier.

\begin{table*}[htbp!]
\centering
\caption{Results on Kinetics-Sounds, CREMA-D, and AVE. 
Acc$_m$: multimodal accuracy, Acc$_a$: audio accuracy, Acc$_v$: video accuracy.
Optimal results are shown in \textbf{bold}. The suboptimal results are indicated by the \underline{underline}.}
\label{tab:main_results}
\resizebox{0.93\textwidth}{!}{
\begin{tabular}{l|cccc|cccc|cccc}
\toprule
\multirow{2}{*}{\textbf{Method}} &
\multicolumn{4}{c}{\textbf{Kinetics-Sounds}} &
\multicolumn{4}{c}{\textbf{CREMA-D}} &
\multicolumn{4}{c}{\textbf{AVE}} \\
\cmidrule(lr){2-5} \cmidrule(lr){6-9} \cmidrule(lr){10-13}
& Acc$_m$ & Acc$_a$ & Acc$_v$ & Avg
& Acc$_m$ & Acc$_a$ & Acc$_v$ & Avg
& Acc$_m$ & Acc$_a$ & Acc$_v$ & Avg \\
\midrule
Joint training          & 64.61 & 52.03 & 35.47 & 50.70
                       & 70.83 & 61.96 & 38.58 & 57.12
                       & 69.65 & 63.93 & 24.63 & 52.74   \\
MSLR          & 65.91 & 50.92 & 42.30 & 53.04
                       & 71.51 & 63.04 & 41.13 & 58.56
                       & 68.91 & 61.19 & 24.63 & 51.58   \\
G-Blending    & 68.90 & 52.11 & 41.35 & 54.12
                       & 73.41 & 62.42 & 65.37 & 67.07
                       & 71.80 & 60.64 & 39.51 & 57.32   \\
OGM-GE        & 66.79 & 51.09 & 37.86 & 51.91
                       & 71.14 & 61.29 & 39.27 & 57.23
                       & 69.12 & 62.45 & 27.39 & 52.99   \\
Greedy       & 65.32 & 50.58 & 35.97 & 50.62
                       & 69.31 & 62.49 & 38.23 & 56.68
                       & 69.66 & 60.76 & 38.70 & 56.37   \\
PMR          & 65.70 & 52.47 & 34.52 & 50.90
                       & 75.54 & 63.04 & 71.24 & 69.27
                       & 70.89 & 63.18 & 35.57 & 56.55   \\
AGM          & 66.17 & 51.31 & 34.83 & 50.77
                       & 77.86 & 63.34 & 37.54 & 59.58
                       & 71.04 & 62.44 & 40.96 & 58.15   \\
MMPareto     & 70.13 & 56.40 & 53.05 & 59.86
                       & 78.53 & \underline{67.38} & 70.26 & 72.06
                       & 75.81 & 64.34 & 45.39 & 61.85   \\
GGDM        & \underline{75.92} & \textbf{61.41} & \underline{59.01} & \underline{65.45}
                       & \textbf{87.10} & 66.83 & \underline{79.97} & \underline{77.97}
                       & \underline{77.10} & \underline{66.34} & \underline{46.64} & \underline{63.36}   \\
\midrule
IIBalance (Ours)       & \textbf{76.04} & \underline{61.32} & \textbf{60.17} & \textbf{65.84}
                       & \underline{86.45} & \textbf{68.53} & \textbf{80.55 }& \textbf{78.51}
                       & \textbf{78.23} & \textbf{67.50} & \textbf{48.11} & \textbf{64.61}   \\
\bottomrule
\end{tabular}}
\end{table*}

\subsection{Overall Training Objective and Dynamic Scheduling}
We optimize \textit{Stage~I} and \textit{Stage~II} with a lightweight curriculum that prioritizes representation/alignment at early epochs and gradually shifts focus to fusion refinement. Let $t$ denote the current epoch and $T$ the total number of epochs. We adopt a linear annealing schedule defined as:
\begin{equation}
\lambda(t)=\lambda_{\mathrm{start}}\Big(1-\frac{t}{T}\Big),
\end{equation}
where $\lambda_{\mathrm{start}}\in[0,1]$ sets the initial emphasis on Stage I. This design stabilizes training by first learning discriminative unimodal features and prototype-guided alignment, then fine-tuning the uncertainty-aware fusion once the representations become reliable.
The overall objective is expressed as:
\begin{equation}\label{eq:finalloss}
\mathcal{L}=\lambda(t)\mathcal{L}_{\mathrm{I}}+\big(1-\lambda(t)\big)\mathcal{L}_{\mathrm{II}},
\end{equation}
During inference, we use the learned encoders and gating network to compute $\tilde{w}_{m}^{(i)}$, form $Z_{\mathrm{fused}}^{(i)}$, and predict with the fused classifier $f_{\mathrm{fuse}}(Z_{\mathrm{fused}}^{(i)})$.

\section{Experiments}
\subsection{Benchmarks}
We evaluate IIBalance on three standard multimodal benchmarks: Kinetics-Sounds~\cite{kay2017kinetics}, a large-scale audiovisual action dataset with about 19K clips from 34 classes; CREMA-D~\cite{cao2014crema}, an audio-visual emotion dataset with 7,442 clips from 91 actors; and AVE~\cite{tian2018ave}, which contains 4,143 aligned audiovisual segments across 28 events and is commonly used to evaluate sample-adaptive fusion. We adopt accuracy as the evaluation metric. To evaluate the impact of input in different modalities on results, we introduced multimodal accuracy $\mathrm{Acc}_m$ for multimodal inputs, alongside unimodal accuracy $\mathrm{Acc}_a$ and $\mathrm{Acc}_v$ to evaluate audio and video inputs, respectively.

\subsection{Baselines}
We compare IIBalance with competitive multimodal balancing and fusion baselines that cover three mainstream directions. For optimizer- and gradient-based balancing, we selected MSLR~\cite{yao2022modality}, G-Blending~\cite{wang2020what}, OGM-GE~\cite{peng2022balanced}, AGM~\cite{li2023agm}, greedy learning mitigation~\cite{wu2022characterizing}, and GGDM~\cite{hu2025ggdm} for comparison. For prototype-guided rebalancing, we compare IIBalance with PMR~\cite{fan2023pmr}, which is most relevant to our prototype alignment. For multi-objective unimodal assistance, we adopted MMPareto~\cite{wei2024mmpareto}. Overall, these baselines span optimization control, structured prototype guidance, and objective coordination,  providing a compact and comprehensive evaluation.

\subsection{Implementation Details}
We follow the audiovisual preprocessing approach adopted in previous studies~\cite{arandjelovic2017look}. Visual frames are sampled at 16\,fps and resized to $224\times224$. Audio signals are resampled to 16\,kHz and converted into log-Mel spectrograms with a 25\,ms window and a 10\,ms hop. For Kinetics-Sounds and AVE, we use the ResNet-18~\cite{he2016resnet} as the visual backbone and a 1D CNN audio encoder following common settings in audiovisual research~\cite{wang2020what}. 
For CREMA-D, we adopt MobileNetV2~\cite{sandler2018mobilenetv2} for face-centric visual inputs and an RNN-based audio encoder~\cite{cao2014crema}. 
We train for 50 epochs using Adam with a learning rate of $1\!\times\!10^{-4}$ and batch size 32. 
The IIB prior $\boldsymbol{\beta}$ is estimated offline from unimodal performances and kept fixed during training.
We set the temperature $\tau=0.07$ and the prototype similarity temperature $\tau_p=0.5$.

\subsection{Main Results}
As shown in Table~\ref{tab:main_results}, IIBalance achieves the best performance across three benchmarks, outperforming both classical balancing strategies and other competitive baselines. Our method yields clear gains in video-only accuracy while keeping audio-only performance competitive, indicating that it effectively boosts the weaker modality instead of overfitting it to residual noise. At the same time, the overall accuracy is improved on every benchmark, demonstrating that imposing an intrinsic information budget and performing sample-aware balancing leads to more effective multimodal fusion than heuristic gradient or loss re-weighting schemes.

\subsection{Effectiveness of Different Components}
\begin{table}[htbp!]
    \centering
    \caption{Ablation of different components of IIBalance. We report $\mathrm{Acc}_m$ as the evaluation metric.}
    \label{tab:ablation_modules}
    \resizebox{0.9\linewidth}{!}{
    \begin{tabular}{lccc}
        \toprule
        Method & Kinetics-Sounds & CREMA-D & AVE \\
        \midrule
        IIBalance        &  \textbf{76.04} & \textbf{86.45} & \textbf{78.23} \\
        \hspace{1em}w/o IIB prior         &  75.20          & 85.10          & 77.10          \\
        \hspace{1em}w/o Stage I (PRA)     &  73.30          & 82.40          & 74.80          \\
        \hspace{1em}w/o Stage II (fusion) &  74.40          & 83.70          & 75.60          \\
        \bottomrule
    \end{tabular}}
\end{table}

Table~\ref{tab:ablation_modules} presents the contribution of the components of IIBalance. Removing the IIB prior and using a uniform budget leads to a consistent but moderate drop on all three datasets, indicating that the dataset-level intrinsic capacity estimate provides a useful prior even when the architecture is unchanged. Disabling \textit{Stage~I} and training the encoders only with unimodal cross-entropy causes the largest degradation, especially on CREMA-D and AVE, which confirms that prototype-guided relative alignment is crucial for preventing dominant modalities from suppressing weaker ones. Removing \textit{Stage~II} and replacing the uncertainty-aware fusion with a fixed fusion rule also harms performance, showing that dynamic, sample-level weighting further refines the benefit of balanced representations.

\subsection{IIB Prior vs Averaged Fusion Weights}
\begin{figure}[htbp!]
    \centering
    \includegraphics[width=0.8\linewidth]{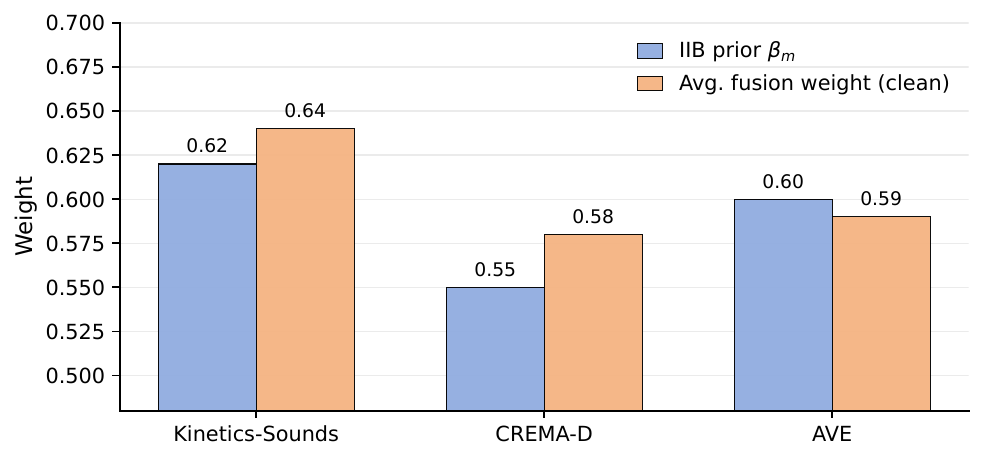}
    \caption{Comparison between the IIB prior $\beta_m$ and the averaged fusion weights on Kinetics-Sounds, CREMA-D, and AVE.}
    \label{fig:iib_prior_weight}
\end{figure}

As shown in Fig.~\ref{fig:iib_prior_weight}, we compare the intrinsic information budget prior with the empirically averaged fusion weights. Across all datasets, the average fusion weights closely follow the IIB prior, indicating that the learned fusion mechanism does not arbitrarily override the dataset-level modality capacity estimated by IIB. Instead, the prior provides a calibrated baseline that reflects the intrinsic discriminative strength of each modality. The small but consistent deviations between the prior and the averaged fusion weights suggest that the uncertainty-aware fusion performs mild sample-level adjustments even in the absence of explicit modality degradation. 

\subsection{Hyperparameter Sensitivity}

\begin{figure}[htbp!]
    \centering
    \begin{subfigure}{0.48\linewidth}
        \centering
        \includegraphics[width=\linewidth]{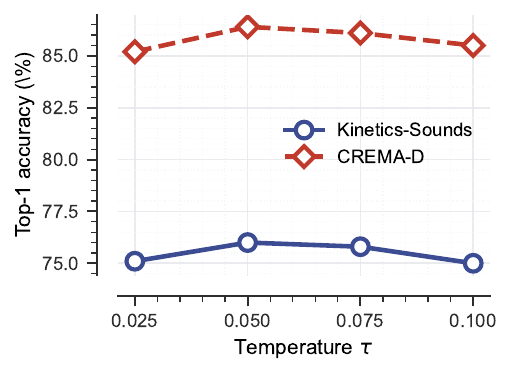}
        \caption{Effect of temperature $\tau$.}
        \label{fig:hyper_tau}
    \end{subfigure}
    \hfill
    \begin{subfigure}{0.48\linewidth}
        \centering
        \includegraphics[width=\linewidth]{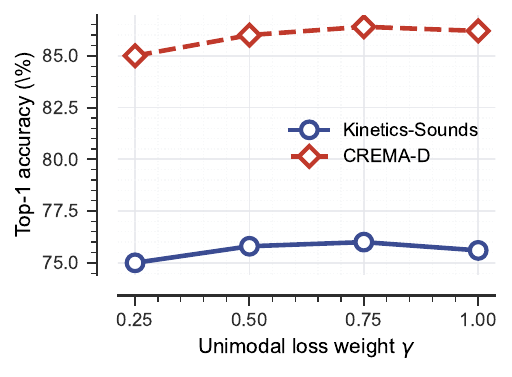}
        \caption{Effect of loss weight $\gamma$.}
        \label{fig:hyper_gamma}
    \end{subfigure}
    \caption{Hyperparameter sensitivity of IIBalance.}
    \label{fig:hyper_sensitivity}
\end{figure}

We analyze the sensitivity of IIBalance to the temperature parameter $\tau$ and the unimodal-loss weight $\gamma$, depicted in Fig.~\ref{fig:hyper_sensitivity}. We found that IIBalance is insensitive to $\tau$, with a broad optimum at moderate values and only mild drops at extremes. For $\gamma$, any non-zero weight consistently improves over removing unimodal regularization, and performance varies smoothly with a shallow best region at intermediate values. The curves on both benchmarks show consistent trends, suggesting that IIBalance does not rely on heavy hyperparameter tuning.

\section{Conclusion}
In this work, we presented IIBalance, a relatively balanced multimodal learning framework that addresses the fundamental challenge of modality dominance in multimodal systems.
Extensive experiments demonstrate that our method consistently improves performance under imbalanced conditions, outperforming existing balanced fusion techniques. We primarily validated its effectiveness based on audio–visual settings. In the future, we plan to learn and update budgets online with self-supervision and extend the framework to broader multimodal scenarios beyond audio–visual recognition.

\bibliographystyle{IEEEbib}
\bibliography{cite}

@inproceedings{kay2017kinetics,
  title={The Kinetics Human Action Video Dataset},
  author={Kay, Will and Carreira, Joao and Simonyan, Karen and et al.},
  booktitle={arXiv preprint arXiv:1705.06950},
  year={2017}
}

@inproceedings{arandjelovic2017look,
  title={Look, Listen and Learn},
  author={Arandjelovic, Relja and Zisserman, Andrew},
  booktitle={Proceedings of the IEEE International Conference on Computer Vision},
  year={2017}
}

@article{cao2014crema,
  title={CREMA-D: Crowd-sourced Emotional Multimodal Actors Dataset},
  author={Cao, Huimin and Cooper, David G and Keutmann, Michael K and Gur, Ruben C and et al.},
  journal={IEEE Transactions on Affective Computing},
  year={2014}
}

@inproceedings{tian2018ave,
  title={Audio-Visual Event Localization in Unconstrained Videos},
  author={Tian, Yapeng and Shi, Jing and Li, Bochen and et al.},
  booktitle={Proceedings of the European Conference on Computer Vision},
  year={2018}
}

@inproceedings{he2016resnet,
  title={Deep Residual Learning for Image Recognition},
  author={He, Kaiming and Zhang, Xiangyu and Ren, Shaoqing and et al.},
  booktitle={Proceedings of the IEEE/CVF Conference on Computer Vision and Pattern Recognition},
  year={2016}
}

@inproceedings{sandler2018mobilenetv2,
  title={MobileNetV2: Inverted Residuals and Linear Bottlenecks},
  author={Sandler, Mark and Howard, Andrew and Zhu, Menglong and et al.},
  booktitle={Proceedings of the IEEE/CVF Conference on Computer Vision and Pattern Recognition},
  year={2018}
}

@article{geirhos2020shortcut,
  title   = {Shortcut Learning in Deep Neural Networks},
  author  = {Geirhos, Robert and Jacobsen, J{\"o}rn-Henrik and Michaelis, Claudio and et al.},
  journal = {Nature Machine Intelligence},
  year    = {2020}
}

@inproceedings{subedar2019uncertainty,
  title     = {Uncertainty-Aware Audiovisual Activity Recognition Using Deep Bayesian Variational Inference},
  author    = {Subedar, Mahesh and Krishnan, Ranganath and  Meyer, Paulo-L{\'o}pez and et al.},
  booktitle = {Proceedings of the IEEE/CVF International Conference on Computer Vision},
  year      = {2019}
}

@inproceedings{tian2020uno,
  title     = {UNO: Uncertainty-Aware Noisy-Or Multimodal Fusion for Unanticipated Input Degradation},
  author    = {Tian, Junjiao and Cheung, Wesley and Glaser, Nathaniel and et al.},
  booktitle = {Proceedings of the IEEE International Conference on Robotics and Automation},
  year      = {2020}
}

@inproceedings{yao2022modality,
  title     = {Modality-specific Learning Rates for Effective Multimodal Additive Late-fusion},
  author    = {Yao, Yiqun and Mihalcea, Rada},
  booktitle = {Findings of the Association for Computational Linguistics: ACL 2022},
  year      = {2022},
}

@inproceedings{wang2020what,
  title     = {What Makes Training Multi-Modal Classification Networks Hard?},
  author    = {Wang, Weiyao and Tran, Du and Feiszli, Matt},
  booktitle = {Proceedings of the IEEE/CVF Conference on Computer Vision and Pattern Recognition},
  year      = {2020}
}

@inproceedings{peng2022balanced,
  title     = {Balanced Multimodal Learning via On-the-fly Gradient Modulation},
  author    = {Peng, Xiaokang and Wei, Yake and et al.},
  booktitle = {Proceedings of the IEEE/CVF Conference on Computer Vision and Pattern Recognition},
  year      = {2022}
}

@inproceedings{wu2022characterizing,
  title     = {Characterizing and Overcoming the Greedy Nature of Learning in Multi-modal Deep Neural Networks},
  author    = {Wu, Nan and Jastrz{\k{e}}bski, Stanis{\l}aw and Cho, Kyunghyun and et al.},
  booktitle = {Proceedings of the 39th International Conference on Machine Learning},
  year      = {2022},
}

@inproceedings{fan2023pmr,
  title     = {{PMR}: Prototypical Modal Rebalance for Multimodal Learning},
  author    = {Fan, Yunfeng and Xu, Wenchao and Wang, Haozhao and et al.},
  booktitle = {Proceedings of the IEEE/CVF Conference on Computer Vision and Pattern Recognition},
  year      = {2023}
}

@inproceedings{li2023agm,
  title     = {Boosting Multi-modal Model Performance with Adaptive Gradient Modulation},
  author    = {Li, Hong and Li, Xingyu and Hu, Pengbo and et al.},
  booktitle = {Proceedings of the IEEE/CVF International Conference on Computer Vision},
  year      = {2023}
}

@inproceedings{wei2024mmpareto,
  title     = {{MMPareto}: Boosting Multimodal Learning with Innocent Unimodal Assistance},
  author    = {Wei, Yake and Hu, Di},
  booktitle = {Proceedings of the 41st International Conference on Machine Learning},
  year      = {2024},
}

@inproceedings{hu2025ggdm,
  title     = {Geometric Gradient Divergence Modulation for Imbalanced Multimodal Learning},
  author    = {Hu, Disen and Jiang, Xun and Sun, Zhe and et al.},
  booktitle = {Proceedings of the 33rd ACM International Conference on Multimedia},
  year      = {2025},
}

@inproceedings{lee2023multimodal,
  title={Multimodal Prompting with Missing Modalities for Visual Recognition},
  author={Lee, Yi-Chun and Tsai, Yi-Hsuan and Chiu, Wei-Chen and et al.},
  booktitle={IEEE/CVF Conference on Computer Vision and Pattern Recognition},
  year={2023}
}

@inproceedings{han2023trusted,
  title={Trusted Multi-View Classification with Dynamic Evidential Fusion},
  author={Han, Zongbo and Zhang, Changqing and Fu, Huazhu and et al.},
  booktitle={IEEE transactions on pattern analysis and machine intelligence},
  year={2023}
}

@InProceedings{pmlr-v162-huang22e,
  title     = {Modality Competition: What Makes Joint Training of Multi-modal Network Fail in Deep Learning? ({P}rovably)},
  author    = {Huang, Yu and Lin, Junyang and Zhou, Chang and et al.},
  booktitle = {Proceedings of the 39th International Conference on Machine Learning},
  year      = {2022},
}

@inproceedings{han2022mmdynamics,
  author    = {Han, Zongbo and Yang, Fan and Huang, Junzhou and et al.},
  title     = {Multimodal Dynamics: Dynamical Fusion for Trustworthy Multimodal Classification},
  booktitle = {Proceedings of the IEEE/CVF Conference on Computer Vision and Pattern Recognition},
  year      = {2022},
}

@inproceedings{zhang2024mla,
  author    = {Zhang, Xiaohui and Yoon, Jaehong and Bansal, Mohit and et al.},
  title     = {Multimodal Representation Learning by Alternating Unimodal Adaptation},
  booktitle = {Proceedings of the IEEE/CVF Conference on Computer Vision and Pattern Recognition},
  year      = {2024},
}

@inproceedings{wei2024valuation,
  author    = {Wei, Yake and Feng, Ruoxuan and Wang, Zihe and et al.},
  title     = {Enhancing Multimodal Cooperation via Sample-level Modality Valuation},
  booktitle = {Proceedings of the IEEE/CVF Conference on Computer Vision and Pattern Recognition},
  year      = {2024},
}

@inproceedings{zhang2023qmf,
  author    = {Zhang, Qingyang and Wu, Haitao and Zhang, Changqing and et al.},
  title     = {Provable Dynamic Fusion for Low-Quality Multimodal Data},
  booktitle = {Proceedings of the 40th International Conference on Machine Learning},
  year      = {2023},
}

@article{xu2023mmcosine,
  author  = {Xu, Ruize and Feng, Ruoxuan and Zhang, Shi-Xiong and et al.},
  title   = {MMCosine: Multi-Modal Cosine Loss Towards Balanced Audio-Visual Fine-Grained Learning},
  journal = {IEEE International Conference on Acoustics, Speech and Signal Processing},
  year    = {2023},
}

\end{document}